\documentclass[twocolumn,prl,showpacs,amsmath,amssymb]{revtex4-1}
\usepackage{epsfig}
\usepackage{amsmath}
\usepackage{psfrag}
\usepackage{units}
\usepackage{graphicx,natbib,amssymb}

\def\bbrel#1#2#3{\mathrel{\mathop{\kern0pt #2}\limits^{#1}_{#3}}}
\usepackage{amsfonts,latexsym,cancel} 

\DeclareGraphicsExtensions{.pdf,.png,.jpg}
\usepackage{subfigure}
\usepackage{multirow}
\begin{document}

\title{Fluctuations, correlations and the estimation of concentrations inside cells}

\author{Emiliano P\'erez Ipi\~na and Silvina Ponce Dawson}
\affiliation{Departamento de
  F\'\i sica, FCEN-UBA, and IFIBA, CONICET,\\
    Ciudad Universitaria,
  Pabell\'on I, (1428) Buenos Aires, Argentina}

% Revision date - uncomment to exclude date in the final version
\date{\today}

% 200 words max Abstract
\begin{abstract}
Information transmission in cells occurs quite accurately even
when concentration changes are ``read'' by individual target
molecules.  In this Letter we study molecule number fluctuations when
molecules diffuse and react.  We show that, for immobile binding
sites, fluctuations in the number of bound molecules are averaged out
on a relatively fast timescale due to correlations. This result can
explain the observed co-existence of highly fluctuating instantaneous
transcriptional activities and of relatively stable protein
concentrations shortly after the beginning of transcription.  We also
show that bound molecule numbers fluctuate with one or two
characteristic timescales depending on the concentration of free
molecules. This transition can explain changes in enzyme activity
observed at the single molecule level.

\end{abstract}

%\emph{Key words:} Diffusion; Reactions; fluorescence; fluctuations}

\maketitle
% New page
\clearpage

The transmission of information in cells usually involves changes in
concentration that are ``read'' by target molecules.  This occurs in a
fluctuating environment. Yet cells respond quite reliably to various
changes~\cite{Bialek19072005,PhysRevLett.100.258101}. The accuracy of
the reading mechanism is key in the case of morphogens, molecules
whose non-uniform distribution results in cell
differentiation~\cite{Bialek:Cell2007}. Most often this patterning
process involves the binding of transcription factors to sites on DNA
controlling the levels of proteins production. The relationship
between the concentration of a protein and of the transcription factor
that regulates its production depends on various binding
processes. How faithful the spatial distribution of protein
concentration reflects that of the transcription factor depends on how
the concentration of the latter is read by the binding sites.  This
relationship has been studied during the early stages of development
of {\it Drosophila melanogaster} embryos in great detail. The analysis
of the variability of the concentrations of the protein Hunchback (Hb)
and of the transcription factor Bicoid (Bcd) involved in its
production shows that the resulting pattern is compatible with
detecting [Bcd] with a 10\% error~\cite{Bialek:Cell2007}. Considering
the random arrivals of individual Bcd molecules to a small
neighborhood around a putative DNA binding site the calculations
of~\cite{Bialek:Cell2007} concluded that only after a long time ($\sim
2 h$) compared to the developmental time of the embryo [Bcd] could be
inferred with this level of precision.  In~\cite{Bialek:Cell2007} a
spatial averaging between neighboring nuclei was invoked to reconcile
this computation with the observations.  As in~\cite{Berg1977193}, the
time computed in~\cite{Bialek:Cell2007} depends on the diffusion
coefficient of Bcd for which there are estimates that differ by over
an order of magnitude~\cite{Gregor:Cell2007,AbuArish:BJ2010}. These
estimates have recently been shown to be compatible~\cite{Sigaut2014}
if they are assumed to correspond to the two effective diffusion
coefficients that describe the transport of molecules that diffuse and
react~\cite{Pando2006}.  In fact, Bcd, being a transcription factor,
diffuses and reacts at least with putative binding sites on DNA. The
analysis of~\cite{Sigaut2014} shows that the two effective
coefficients of Bcd can be very different between themselves.  What is
the diffusion coefficient that sets the limit for the
precision with which [Bcd] can be read? Protein production, on the
other hand, is determined by the number of bound rather than free
transcription factor molecules. How do fluctuations in the number of
bound and free molecules relate to each other? In this paper we
address these two issues. Building on previous works on the analysis
of optical experiments when molecules diffuse and
react~\cite{Ipina2014,Ipina2013,Sigaut2010} we derive expressions for
the relative error with which the concentrations involved can be
estimated as a function of the observation time and apply it to
interpret recent results on the variability of mRNA
production~\cite{Little2013789,Gregor2014364} and on enzyme turnovers
at the single molecule level~\cite{English2006}.  In particular, we
show that the interaction with immobile binding sites introduces
correlations that reduce the variance of the bound molecules number
and that the observation time that is needed to estimate the
concentration of these molecules with a given accuracy depends on a
relatively fast correlation time.

In this Letter we consider a system of particles ({\it e.g.},
transcription factors), $P^{(f)}$, that diffuse
with (free) coefficient, $D_f$, and react with  binding
sites, $S$, according to~\cite{Pando2006,Sigaut2010,Ipina2013}:
\begin{equation}\label{eq:schemeP}
P^{(f)} + S\,\,\raisebox{-2.5ex}{$\stackrel{\stackrel{\textstyle k_{on}}
{\textstyle{\longrightarrow }}}{\stackrel{\textstyle{\longleftarrow}}
{\textstyle{k_{off}}}}$} \,\, P^{(b)} .
\end{equation}
  We assume that the binding sites diffuse with coefficient $D_S\ll
  D_f$ ($D_S$ can be zero) and that the mass of $S$ is so large that
  the free coefficient of $P^{(b)}$ is $D_S$ too. We consider a total
  volume, $V_T$, over which the molecules diffuse and the
  concentrations, $[P^{(f)}]$, $[P^{(b)}]$, $[S]$, are approximately
  constant, uniform and in equilibrium among themselves
  ($[P^{(f)}][S]=K_D [P^{(b)}]$), and an observation volume,
  $V_{obs}$, where we count the number of molecules of the three
  species, $N^{(f)}$, $N^{(b)}$ and $N^{(S)}$, every time step,
  $dt$. These are the stochastic variables of the problem which means
  satisfy $\langle N^{(f)}\rangle=[P^{(f)}]V_{obs}$, $\langle
  N^{(b)}\rangle=[P^{(b)}]V_{obs}$ and $\langle
  N^{(S)}\rangle=[S]V_{obs}$ if $D_s\neq 0$.  If $D_S=0$ and
  $V_{obs}\ll V_T$, there could be a local equilibrium in $V_{obs}$
  slightly different from the one in $V_T$ that depends on the (fixed)
  total number of binding sites in $V_{obs}$, $N_{ST}\equiv
  N^{(b)}+N^{(S)}$. The aim is to determine the difference between the
  mean and the average, $\overline{N}^{(s)}(T_{obs})=\frac{\sum_l
    N_\ell^{(s)}}{n}$ of each stochastic variable ($s=f,b,S$) after an
  observation time, $T_{obs}=n dt$ ({\it i.e.}, from a sequence
  $\{N_\ell^{(s)} \equiv N^{(s)}(t_\ell)\}_{\ell=0}^{n-1}$). This
  difference will allow us to estimate the time that is needed to
  derive $\langle N^{(s)}\rangle$ with a given accuracy from counting
  molecules in $V_{obs}$. The (mean) square difference between
  $\overline{N}^{(s)}(T_{obs})$ and $\langle N^{(s)}\rangle$ is the
  variance of the average,
\begin{eqnarray}
&&\operatorname{var}\left(\overline{N}^{(s)}(T_{obs})\right)=\operatorname{var}\left(\frac{1}{n}\sum_{\ell=0}^{n-1}
N_\ell^{(s)}\right)\nonumber\\&&=\frac{1}{n^2}\langle\sum_{\ell,k=0}^{n-1}\left(N_\ell^{(s)}-\langle
N^{(s)}\rangle\right)\left(N_k^{(s)}-\langle N^{(s)}\rangle\right)\rangle ,
\label{eq:var}
\end{eqnarray}
which is related to the autocorrelation function (ACF) of the particle
number fluctuations.  Using the normalization of FCS experiments, the
ACF for species $s$ is given by $G^{(s)}(\tau)/\langle
N^{(s)}\rangle^2$ with $G^{(s)}(\tau=j dt) = \lim_{n\rightarrow\infty}
\sum_{\ell=0}^{n-1} \left(N_\ell^{(s)}-\langle N^{(s)}\rangle\right)
\left(N_{\ell+j}^{(s)}-\langle N^{(s)}\rangle\right) /n$. For systems
with one species that diffuses with coefficient, $D$, and a Gaussian
$V_{obs}$ of width, $w_r$, it is~\cite{Krichevsky2002}:
\begin{equation}
G(\tau) = 
\frac{\operatorname{var}(N)}{
  \left(1+\frac{\tau}{\tau_D}\right)^{3/2}} ,\label{eq:free_diff}
\end{equation}
with $\tau_D={w_r^2}/{4D}$, $N$ the number of molecules in $V_{obs}$ 
and
${\operatorname{var}(N)}=\langle N\rangle$.  $G(\tau)$ in Eq.~(\ref{eq:free_diff})
is relatively flat with $G(\tau)\approx
\operatorname{var}(N)$ for $\tau\le \tau_D$ and
$G(\tau)\approx 0$ otherwise. We then approximate
$\frac{1}{n}\sum_{\ell=0}^{n-1}\left(N_\ell-\langle
N\rangle\right)\left(N_{k}-\langle N\rangle\right)\approx
{\operatorname{var}(N)}$ for $\vert k-\ell\vert\le \tau_D/dt$ and
assume it is negligible otherwise. If $n\ge \tau_D/dt$ we then obtain:
\begin{equation}
\operatorname{var}\left(\overline{N}(T_{obs})\right)\approx
\frac{1}{n}\sum_{k=-\tau_D/dt}^{\tau_D/dt}
\operatorname{var}(N) \approx \frac{2\tau_D}{T_{obs}}
\operatorname{var}(N),
\label{eq:error_av}
\end{equation}
with the last approximation being valid for $\tau_D\gg dt$. If $n\le
\tau_D/dt$ the same formula is obtained but with 1 instead of
$\tau_D/T_{obs}$. Replacing $\tau_D={w_r^2}/{4D}$ in
Eq.~(\ref{eq:error_av}) we obtain a similar error of the average as
the one considered in~\cite{Bialek:Cell2007,Berg1977193}.
Eq.~(\ref{eq:error_av}) implies that the relative error,
$\Delta_r(\overline{N})\equiv(\operatorname{var}\left(\overline{N}(T_{obs})\right))^{1/2}/
\langle N\rangle$ decreases with the correlation time and
$\operatorname{var}(N)$. The necessary time to obtain an estimate with
relative error $\alpha$, on the other hand, is $T_{obs}(\alpha)\sim
\tau_D \operatorname{var}(N)/(\alpha \langle N\rangle)^2$.

When the reaction-diffusion system of species that corresponds to
Eq.~(\ref{eq:schemeP}) is considered there is more than one
correlation time. Working in Fourier space as in~\cite{Krichevsky2002}
we obtain 3 branches of eigenvalues, $\lambda_i$, that rule the
fluctuations dynamics.  $\lambda_1$ always corresponds to the free
diffusion time of the binding sites, $\tau_S\equiv
w_r^2/4D_S$~\cite{Sigaut2010,Ipina2013}. $\lambda_2$ and $\lambda_3$
have a clear meaning in the fast diffusion ($\tau_f\equiv w_r^2/(4
D_f)\ll \tau_r\equiv {1}/({k_{off}(1+[P^{(f)}]/K_D+[S]/K_D)})$), and
in the fast reaction ($\tau_r\ll\tau_f$) limits. In both limits the
eigenvalues can be written as $-\nu_i-D_i q^2$, with $q$ the variable
conjugate to position in Fourier space. They contribute to
$G^{(s)}(\tau)$ with an additive term of the form:
\begin{equation}
G_i^{(s)}(\tau) = \frac{{G}_{oi}^{(s)}}{\left(1 +
\frac{\tau}{\tau_{Di}}\right)^{3/2}} e^{-\nu_i \tau},\quad\,  s=f,b,S,\label{eq:G_i}
\end{equation}
where $\tau_{Di}={w_r^2}/{4D_i}$ for a Gaussian $V_{obs}$ and the
weights, ${G}_{oi}^{(s)}$, are linear combinations of the covariances
between the stochastic variables that satisfy $G_o^{(s)}\equiv\sum_i
{G}_{oi}^{(s)}={\operatorname{var}(N^{(s)})}$~\cite{Ipina2014,support_info}.
In both limits it is $\nu_i=0$ if $D_i\neq 0$ and viceversa. Thus,
there is a single correlation time, $\tau_i$, associated to each
eigenvalue and to each term, $G_i^{(s)}(\tau)$, which is diffusive (if
$D_i\neq 0$) or determined by the reactions only (if $D_i= 0$).
Ordering the times so that $\tau_1\geq\tau_2\geq\tau_3$, it is
$\tau_1=\tau_S$ always, unless $D_S=0$ in which case it disappears
from the ACF, as explained later. In the fast diffusion limit, it is
$\tau_3=w_r^2/4 D_f$ while $\tau_2= w_r^2/4 D_S$ (if $D_S\neq 0$) or
$\tau_2= \tau_{off}\equiv k_{off}^{-1}\langle N^{(S)}\rangle/N_{ST} $
(if $D_S=0$).  In the fast reaction limit it is $\tau_3=\tau_r$ while
$\tau_2=\tau_{coll}\equiv w_r^2/4D_{coll}$ with $D_{coll}$ a
(reaction-dependent) ``effective'' diffusion coefficient: $D_{coll} =
(D_f+\frac{[S]^2}{K_D [S_T]} D_S) /(1+\frac{[S]^2}{(K_D [S_T])})$
where $[S_T]=[P^{(b)}]+[S]$~\cite{Pando2006,Sigaut2010}.
$\tau_{coll}$ can be of the order of $\tau_f$ even if a relatively
large fraction of the particles is bound~\cite{Pando2006}.  As for Eq.~(\ref{eq:free_diff}), we assume that
$G_i^{(s)}(\tau)\approx {G}_{oi}^{(s)}$ if $\vert\tau\vert\leq\tau_i$
and negligible otherwise. Applying this approximation to compute
$\operatorname{var}\left(\overline{N}^{(s)}\right)$ and assuming
$\tau_i\gg dt$ and $n\ge \tau_i/dt$ we obtain:
\begin{equation}
    (\Delta_r
  \overline{N}^{(s)})^2(T_{obs})=\frac{\operatorname{var}\left(\overline{N}^{(s)}\right)}{\langle
    N^{(s)}\rangle^2}\approx
           \frac{{\operatorname{var}(N^{(s)})}}{\langle
    N^{(s)}\rangle^2}\sum_{i=1}^3W_i^{(s)}\frac{2\tau_{i}}{T_{obs}},
      \label{eq:var_many}
  \end{equation}
where $W_i^{(s)}\equiv {G_{oi}^{(s)}}/{G_o^{(s)}}$ (so that
$\sum_iW_i^{(s)}=1$).  Also in this case the relative errors increase
with the correlation times and $\operatorname{var}(N^{(s)})$ and
decrease when $\langle N^{(s)}\rangle$ increases. As before, the
ratio $\tau_i/T_{obs}$ has to be replaced by 1 in Eq.~(\ref{eq:var_many}) if
$T_{obs}\le\tau_{i}$.

\begin{figure}[H]
   \begin{center}
      \includegraphics*[width=3.25in]{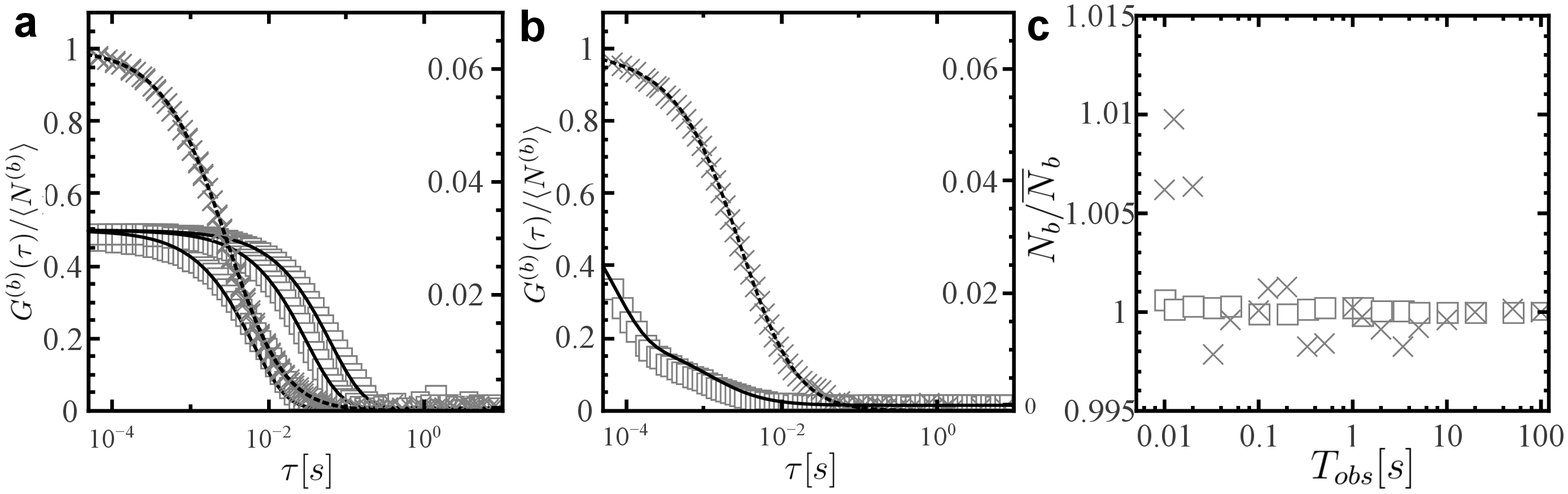}
\caption{(a) $G^{(b)}(\tau)/\langle N^{(b)}\rangle$ for a system of particles that diffuse
  and react with immobible ($\square$) or mobile ($\times$) sites in
  the limit of fast diffusion for $k_{off} = 0.5 s^{-1}$, $1 s^{-1}$,
  $5 s^{-1}$. Variations of $G^{(b)}(\tau)/\langle N^{(b)}\rangle$ with $k_{off}$ are apparent for
  $D_S=0$.  (b) $G^{(b)}(\tau)/\langle N^{(b)}\rangle$ as in (a) but for other
  concentrations, $k_{off}=400 s^{-1}$ so that the fast reaction
    limit holds and $D_S = 0$ ($\square$) or $5 \mu m^2 s^{-1}$
    ($\times$). In (a) and (b), the curves are the theoretical
    predictions for $D_S\neq 0$ (dashed, scale at left) and $D_S=0$
    (solid, scale at right).  (c) $\overline{N^{(b)}}(T_{obs})/\langle
    N^{(b)}\rangle$ {\it vs.}  $T_{obs}$ for the same parameters as in
    (b) and $D_S=0$ ($\square$) or $0.2 \mu m^2 s^{-1}$
    ($\times$).}
      \label{fig:figura1}
   \end{center}
\end{figure}

How relevant each correlation time is for a species, $s$,
depends on the relative weight, $W_i^{(s)}$.  If $D_S\neq 0$, we
assume as usual~\cite{Krichevsky2002} that
$\operatorname{cov}(N^{(s)}_{\ell},N^{(v)}_{\ell})=\langle
N^{(s)}\rangle \delta_{s,v}$, with $s, v=f,b,S$~\cite{support_info}.
If $D_S=0$, given that $N^{(b)}+N^{(S)}$ is fixed, we assume that
$N^{(b)}$ and $N^{(S)}$ are binomial so that
$\operatorname{var}(N^{(S)})=\operatorname{var}(N^{(b)})=\langle
N^{(b)}\rangle \langle N^{(S)}\rangle /N_{ST}$, and
$\operatorname{cov}(N^{(b)}_{\ell},N^{(S)}_{\ell})=-\operatorname{var}(N^{(b)})$.
The weights for $s=f$ do not depend on this assumption, but those of
$s=b,S$, do. Namely, $W_1^{(f)}=0$ always while
$W_1^{(b)}=0$ if $D_S=0$ due to the correlations between $N^{(b)}$ and
$N^{(S)}$ but it is finite ($=f_b\equiv \langle
N^{(b)}\rangle/N_{ST} $)  even if $D_S$ is arbitrarily small~\cite{support_info}. The
finite change in $W_1^{(b)}$ is reflected in the ACF as illustrated in
Figs.~\ref{fig:figura1}~(a,b) where we show $G^{(b)}(\tau)/\langle
N^{(b)}\rangle$ computed from numerically generated time-series
(symbols) and using Eq.~(\ref{eq:G_i}) (lines) with the analytic
weights of the fast diffusion (a) and the fast reaction (b) limits. To
generate the series we performed stochastic simulations of particles
that diffuse with $D_f=19\mu m^2/s$ and react according to
Eq.~(\ref{eq:schemeP}) with $K_D=0.192\mu M$ using a Gillespie-like
scheme~\cite{Gillespie,Ipina2013}.
$G_o^{(b)}$ and $\operatorname{var}(N^{(b)})$ are smaller for the
cases with $D_S=0$ than for the corresponding ones with $D_S\neq 0$
although $\langle N^{(b)}\rangle$ is the same in each subfigure. 
The difference is noticeable because $f_b \approx 0.92$.
Fig.~\ref{fig:figura1}~(a) also illustrates the change of the relevant
timescales that is observed in the fast diffusion limit when $D_S=0$.
 Namely, in this limit, fluctuations in $N^{(b)}$ depend
on $\tau_S$ if $D_S\neq 0$ and on $\tau_{off}$ if $D_S= 0$. In this
figure we have superimposed the results obtained for various values of
$k_{off}$.  Changes of the ACF with $k_{off}$ are unobservable for
$D_S\neq 0$ while they are noticeable for $D_S=0$.  The change in the
relevant timescale in the fast reaction limit is illustrated in
Fig.~\ref{fig:figura1}~(c). There we observe that $\overline{N}^{(b)}$
approaches its expected value, $\langle N^{(b)}\rangle$, faster if
$D_S=0$ than if it is $D_S=0.2 \mu m^2 s^{-1}$ since the slowest
timescale of the latter, $\tau_S$, is absent in the former.

\begin{figure}[H]
   \begin{center}
      \includegraphics*[width=3.25in]{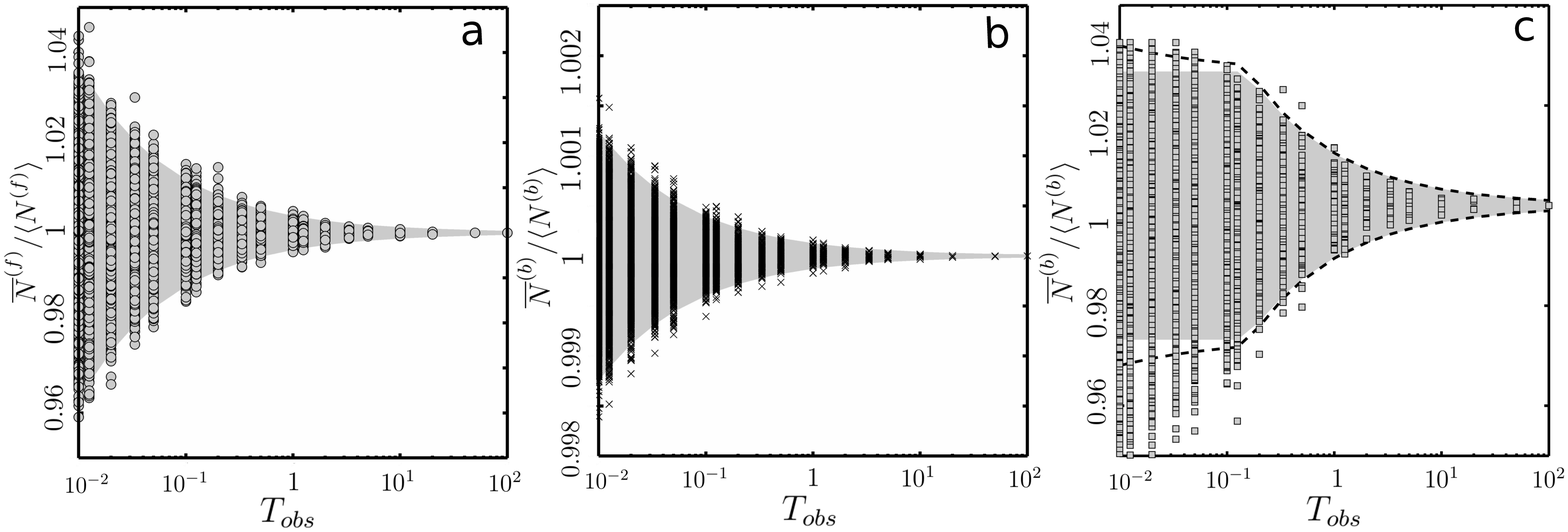}
      \caption{Normalized average numbers of free, $N^{(f)}$ (a), and
        bound, $N^{(b)}$ (b, c), particles obtained from stochastic
        simulations ($\square$) of reaction-diffusion systems with
        $D_S=0$ and relative errors, $\Delta_r(\overline{N}^{(f,b)})$,
        given by Eq.~(\ref{eq:var_many}) (gray shaded areas) as
        functions of $T_{obs}$.  In (a,b) the parameters are the same
        as in Fig.~\ref{fig:figura1}~(c) and in (c) as in
        Figs.~\ref{fig:figura1}~(a,b).  In (c) the error given by
        Eq.~(\ref{eq:var_with_error}) is also shown (dashed curves).
      }
      \label{fig:figura2}
   \end{center}
\end{figure}

We compare now the necessary time, $T_{obs}(\alpha)$, to estimate
$\overline{N}^{(f)}$ and $\overline{N}^{(b)}$ with a relative error,
$\alpha$, when $D_S=0$. Identifying $\overline{N}^{(f)}$ and
$\overline{N}^{(b)}$ with the free and the DNA-bound transcription
factor molecules we can apply this comparison to study the
accuracy of transcription, given that the resulting protein
concentration (accumulated up to $T_{obs}$) depends on
$\sum_{\ell=1}^{T_{obs}/dt}
N^{(b)}_\ell=T_{obs}\overline{N}^{(b)}(T_{obs})$.
Eq.~(\ref{eq:var_many}) implies that $T_{obs}(\alpha)$ scales linearly
with $\operatorname{var}\left({N}^{(s)}\right)/\langle
N^{(s)}\rangle^2$. 
We have discussed that, if $D_S=0$,
$\operatorname{var}\left({N}^{(b)}\right)=\langle N^{(b)}\rangle
(1-f_b)$
while fluctuations in $N^{(f)}$ follow a Poisson distribution.
Regardless of the correlation times, then $\overline{N}^{(b)}$ can be
within a few percent of its expected value if $D_S=0$ and $f_b\approx 1$.  An
interesting situation can be
found in the fast reaction limit. In this limit
$\Delta_r(\overline{N}^{(f)})$ and $\Delta_r(\overline{N}^{(b)})$
depend on $\tau_{coll}$ and $\tau_r$ with weights
$W_2^{(f)}=W_3^{(b)}=(1+\beta)^{-1}$ and
$W_3^{(f)}=W_2^{(b)}=1-W_2^{(f)}$ where $\beta = \langle
N^{(S)}\rangle^2/(K_DN_{ST}V_{obs})$. Regardless of $f_b$, $\beta$ can
be larger or smaller than 1 depending on $\langle
N^{(S)}\rangle/\langle N^{(f)}\rangle$. If $\beta< 1$, it is $W_2^{(f)}=W_3^{(b)}>
W_3^{(f)}=W_2^{(b)}=1-W_2^{(f)}$ so that $\overline{N}^{(b)}$
approaches its expected value over a faster timescale than
$\overline{N}^{(f)}$. Furthermore, even if $f_b\sim 1$,
the time, $\tau_{coll}$, can be of
the same order as the particles free diffusion time, $ \tau_f$. This
implies that there are parameters for which $f_b\sim 1$, so that
$\operatorname{var}\left({N}^{(b)}\right)$ is sensitively reduced with
respect to the Poisson case and, at the same time, the effective
diffusion coefficient, $D_{coll}\sim D_f$, so that the slowest
convergence time of $(\overline{N}^{(b,f)})$ is of the same order as
$\tau_f$.  This combination of parameters
is not just a speculation. Analyzing the FCS experiments
of~\cite{AbuArish:BJ2010} on the diffusion of Bcd in {\it Drosophila
  melanogaster} embryos under the assumption that Bcd diffuses and reacts
following Eq.~(\ref{eq:schemeP}) and that the fast reaction limit
holds we found $\tau_{coll}\sim 0.7\tau_f$, $\beta \sim 0.4$ and
$f_b\sim 0.92-0.98$~\cite{Sigaut2014}. We illustrate in
Fig.~\ref{fig:figura2}~(a,b) how, for the same combination of times and
fraction of bound sites as those deduced in~\cite{Sigaut2014},
$\Delta_r(\overline{N}^{(b)})< \Delta_r(\overline{N}^{(f)})$ for any
$T_{obs}$ by a factor that cannot be accounted for by the difference
between $\langle{N}^{(b)}\rangle$ and $
\langle{N}^{(f)}\rangle$. Namely, $\langle N^{(f)}\rangle\sim 2200$,
$\langle N^{(b)}\rangle\sim 22000$ in this figure so that
$\Delta_r(\overline{N}^{(b)})/ \Delta_r(\overline{N}^{(f)})\sim 0.17
(\langle N^{(f)}\rangle/\langle N^{(b)}\rangle)^{1/2}$ for any
$T_{obs}$.
This figure also shows that
Eq.~(\ref{eq:var_many}) provides relatively good estimates of
$\Delta_r(\overline{N}^{(s)})$.
It is important to note that $\Delta_r(\overline{N}^{(b)})$ depends on
the fraction of bound binding sites in $V_{obs}$, which is a
stochastic variable that we approximate by $\langle
N^{(b)}\rangle/N_{ST}$ in Eq.~(\ref{eq:var_many}).  When many
reactions occur during the diffusion timescale ({\it i.e.}, in the
case of Figs.~\ref{fig:figura2}~(a,b)), most of the time the molecule
numbers are in equilibrium between themselves and $f_b$ is close to
its expected value. Something different can occur in systems with
small $V_{obs}$ where the fast diffusion limit holds. This is the
situation that a binding site ``encounters'' when ``trying'' to infer
the concentration of its ligand as considered
in~\cite{Bialek:Cell2007}. A rough way to treat the stochasticity of
$f_b$ is to assume it has an associated error and propagate it in
$\Delta_r(\overline{N}^{(b)})$.  In this way we obtain $\Delta
f_b=(\operatorname{var}(N^{(f)}))^{1/2}K_DV_{obs}/(K_DV_{obs}+\langle
N^{(f)}\rangle$ and, using $\tau_{off}= (1-f_b)/k_{off}$,
\begin{equation}
\left(\Delta_r(\overline{N}^{(b)})\right)^2\approx
  \frac{(1-f_b)}{\langle N^{(b)}\rangle}\frac{\tau_{off}}{T_{obs}}
  \left(1+2\sqrt{\frac{\tau_f}{T_{obs}\langle N_f\rangle}}\right) ,\label{eq:var_with_error}
\end{equation}
where we have assumed that the fast diffusion holds and $T_{obs}\ge
\tau_i$ ($i={off}, {f}$). If $T_{obs}< \tau_i$, the ratio,
$\tau_{i}/T_{obs}$ must be replaced by 1.  Fig.~\ref{fig:figura2}~(c)
illustrates how $\Delta_r(\overline{N}^{(b)})$ varies with $T_{obs}$
when this effect is relevant. It is the equivalent of
Fig.~\ref{fig:figura2}~(b) for a system in the fast diffusion limit
and where Eq.~(\ref{eq:var_with_error}) is also plotted with dashed
curves.  We observe that Eq.~(\ref{eq:var_with_error}) captures well
the decay of $\Delta_r(\overline{N}^{(b)})$ with $T_{obs}$ which
occurs faster than if the convergence of $f_b$ is not considered.
Eq.~(\ref{eq:var_with_error}) can be used to interpret recent
observations  of transcriptional regulation {\it in
  vivo} in {\it Drosophila melanogaster}
embryos~\cite{Little2013789,Gregor2014364}.  These studies show that
the instantaneous production of the Hb mRNA varies up to 50\%
between loci of transcription while the resulting cytoplasmic mRNA and
protein concentrations in a volume embracing a nucleus
fluctuate by less than 10\%. The protein, Hb, is long-lived and
accumulates with time, so that time averaging can be responsible for
smoothing the instantaneous fluctuations out. The increase in
precision, however, cannot be explained by time averaging so
that the occurrence of some spatial averaging was invoked
in~\cite{Little2013789}. The diffusion of the free transcription
factors (our $P_f$) between loci and its effect on the convergence of
$f_b$ as included in Eq.~(\ref{eq:var_with_error}) could be the
mechanism that underlies the smoothing out of fluctuations in
$\overline{N}^{(b)}$ and thus, on the number of Hb mRNA molecules, on
a faster timescale than the one prescribed by time averaging
(Eq.~(\ref{eq:var_many}).

In this Letter we have presented results obtained in two opposite
limits. From their differences we can infer the types of situations
that may be found in between. The convergence times of the average
number of free and bound paticles are different depending on the
limit.  Various quantities can be varied to change the ratio,
$\tau_f/\tau_r$, that rules the transition between limits. As
expected, $\tau_f$, decreases with increasing $D_f$ and $\tau_r$ with
increasing $k_{off}$~\cite{Ipina2013}. $\tau_f$ also decreases with
$V_{obs}$ and $\tau_r$ with increasing concentrations.  Thus,
by considering a small or a large $V_{obs}$ not only the
fluctuation sizes change due to the different numbers of particles but
also the correlation times change with their corresponding effect on
the errors in molecule number estimates. This has implications in
morphogenesis.
It can also be related to the changes observed in 
single enzyme activities~\cite{English2006}.  The first step of a
Michaelis-Menten scheme in which a substrate, $P^{(f)}$, binds to an
enzyme, $S$, and is then transformed into a product at a rate
proportional to $[P^{(b)}]$ is given by Eq.~(\ref{eq:schemeP}). The
observation of these reactions at the single molecule level
showed a  distribution of waiting times
between individual turnovers that was monoexponential at low $[P^{(f)}]$
and was characterized by several timescales at high 
$[P^{(f)}]$~\cite{English2006}. 
Based on the results presented in this Letter we can interpret this
change in terms of a transition from the fast diffusion to the fast
reaction limit as $[P^{(f)}]$ increases.  This involves a change in
the timescales of the $N^{(b)}$ fluctuations: from the single
non-diffusive timescale, $\tau_{off}$, to two timescales, $\tau_r$ and
$\tau_{coll}$ (with the additional timescale, $\tau_f$, if the effect
of Fig.~\ref{fig:figura2}~(c) is included).  The appearance of a
diffusive timescale as the fast reaction limit is approached can also
underlie the change to a ``broad'' dwell time distribution observed
in~\cite{English2006}.

\section*{Acknowledgments}
This research has been supported by UBA (UBACyT 20020100100064) and
ANPCyT (PICT 2010-1481 and PICT 2010-2767). 

\bibliography{fluctuations_cells_no_url}

\end{document}

\title{Supporting Information for Fluctuations, correlations and the estimation of concentrations inside cells}

\author{Emiliano P\'erez Ipi\~na and Silvina Ponce Dawson}
\affiliation{Departamento de
  F\'\i sica, FCEN-UBA, and IFIBA, CONICET, Buenos Aires, Argentina}

\maketitle
% New page
%\clearpage

\section{The system}
We consider a system~\cite{Pando2006,Sigaut2010, Ipina2013} composed of molecules $P^{(f)}$, that diffuse with free coefficient, $D_f$, and react with binding sites, $S$, according to:
\begin{equation}\label{eq:schemeP}
P^{(f)} + S\,\,\raisebox{-2.5ex}{$\stackrel{\stackrel{\textstyle k_{on}}
{\textstyle{\longrightarrow }}}{\stackrel{\textstyle{\longleftarrow}}
{\textstyle{k_{off}}}}$} \,\, P^{(b)} .
\end{equation}
where the rates of binding and unbinding define the dissociation constant, $K_D=\frac{k_{off}}{k_{on}}$.  We assume that the binding sites are much larger than the particles so that $P^{(b)}$ and $S$ diffuse with $D_S\ll D_f$. We consider a total volume, $V_T$, over which the molecules diffuse and the concentrations, $[P^{(f)}]$, $[P^{(b)}]$, $[S]$, are approximately constant, uniform and in equilibrium among themselves ($[P^{(f)}][S]=K_D [P^{(b)}]$), and an observation volume, $V_{obs}$, where we count the number of  molecules, $N^{(f)}$, $N^{(b)}$ and $N^{(S)}$, every time step, $dt$. We assume that the mean, $\langle N^{(f)}\rangle$, satisfies $\langle N^{(f)}\rangle=[P^{(f)}]V_{obs}$. The relations $\langle N^{(b)}\rangle =[P^{(b)}]V_{obs}$ and $\langle N^{(S)}\rangle=[S]V_{obs}$ (which, due to the equilibrium condition imply that $\langle N^{(b)}\rangle=(\langle N^{(b)}\rangle+\langle N^{(S)}\rangle)[P^{(f)}]/({[P^{(f)}]+K_D})$), are assumed to hold if $D_s\neq 0$. If $D_S=0$ we assume that $\langle N^{(b)}\rangle= N_{ST}{[P^{(f)}]}/({[P^{(f)}]+K_D})$ and $\langle N^{(S)}\rangle=N_{ST}-\langle N^{(b)}\rangle$, with $N_{ST}\equiv N^{(b)}+N^{(S)}$, the (fixed) total number of binding sites in $V_{obs}$.

\section{Autocorrelation Function. Analytic calculations}
We consider the sequences, $\{N_\ell^{(s)} \equiv
N^{(s)}(t_\ell)\}_{\ell=0}^{n-1}$, obtained after an observation time,
$T_{obs}=n dt$ for the three species, $s=f,b,S$, and compute
$G^{(s)}(\tau=j dt) = \lim_{n\rightarrow\infty} \sum_{\ell=0}^{n-1}
\left(N_\ell^{(s)}-\langle N^{(s)}\rangle\right)
\left(N_{\ell+j}^{(s)}-\langle N^{(s)}\rangle\right) /n$ for each of
them. Dividing each of these functions by $\langle N^{(s)}\rangle^2$
we obtain the autocorrelation function (ACF) as defined in
Fluorescence Correlation Spectroscopy (FCS) experiments. In this case,
it is the ACF of the molecule number fluctuations of each species in
$V_{obs}$. This computation is easy to perform in the case of
numerical simulations. For the analytic calculations it is
simpler to work as in the case of FCS experiments. Namely, instead of
adding all the particles of species $(s)$ in $V_{obs}$ to compute
$N^{(s)}$, we add all the particles of species $(s)$ in $V_T$ but with
a Gaussian weight: $N^{(s)}= \int_{V_T} d^3\vec{r}I(\vec{r})c^{(s)}$
where $I(\vec{r})= \exp\left(-2\frac{r^2}{w_r^2}\right)$,
$r=\vert\vec{r}\vert$, $w_r$ is the waist of the Gaussian and
$c^{(s)}=\sum_{i_s} \delta (\vec{r}-\vec{r}_{i_s}(t))$ with the sum
running over all the molecules of species $(s)$ and $\vec{r}_{i_s}(t)$
the location of each of them at time $t$. In this way, it is $V_{obs}=
\int d^3\vec{r}I(\vec{r})=(\pi/2)^{3/2}w_r^3$ and:
\begin{equation}
G^{(s)}(\tau) =\langle\delta N^{(s)}(t) \delta N^{(s)}(t+\tau)\rangle = \frac{1}{T_{obs}}\int d\vec{r} \int d\vec{r'}\int_0^{T_{obs}} dt\,  I(\vec{r})I(\vec{r'}) \delta c^{(s)}(\vec{r},t) \delta c^{(s)}(\vec{r'},t+\tau) ,
\label{Go_generica}
\end{equation}
where $\delta c^{(s)} \equiv c^{(s)}(\vec{r},t) -\langle N^{(s)}
\rangle/V_{obs}$. As done in~\cite{Ipina2014,Krichevsky2002}, for the
analytic computation of $G^{(s)}(\tau) $ we calculate the differences,
$\delta c^{(s)}$, for the 3 species of the system, as the solution of
the reaction-diffusion equations linearized around  equilibrium.
This solution can be written in Fourier space in terms of the
(branches of) eigenvalues and eigenvectors of the linear system so that
$G^{(s)}$ can be expressed as:
\begin{eqnarray}
G^{(s)}(\tau) &=& \int
d\vec{q}\vert\hat{I}(\vec{q})\vert^2\sum_{m}X^{(m)}_{j}\exp(\lambda^{(m)}\tau)(X^{-1}\sigma^2)^{(m)}_{j}
\label{Go_krichevsky}
\end{eqnarray}
where the subscript, $j$, refers to the species ($j=1$ for $s=S$,
$j=2$ for $s=f$ and $j=3$ for $s=b$) and the index, $(m)$, labels the
eigenvalues, $\hat{I}(\vec{q})$ is the Fourier transform of
$I(\vec{r})$ and $\vec{q}$ is the conjugate variable of $\vec{r}$, $X$
is the matrix of eigenvectors, $\lambda^{(m)}$ is the $m$-th
eigenvalue and $\sigma^2$ is the matrix of initial correlations
between the species, $\sigma^2_{ij}=\langle \delta N^{(s)}(t)\delta
N^{(s')}(t) \rangle2^{3/2}/V_{obs}$ with $i,j$ the indices corresponding to
species $s$ and $s'$, respectively. In~\cite{Krichevsky2002} it is
assumed that the initial correlations satisfy $\langle \delta
c^{(s)}(\vec{r},t) \delta c^{(s')}(\vec{r'},t)\rangle = \langle
c^{(s)}\rangle \delta_{ij} \delta(\vec{r}-\vec{r'})$ with
$\delta_{ij}$ the Kronecker delta. This assumption implies that the
correlations are spatially short-ranged, that the number of molecules
of different species in $V_{obs}$ are uncorrelated and that, for each
species, they are Poisson distributed. In deriving
Eq.~(\ref{Go_krichevsky}) we have also assumed that the initial
correlations are short ranged $\langle \delta c^{(s)}(\vec{r},t)
\delta c^{(s')}(\vec{r'},t)\rangle \propto\delta(\vec{r}-\vec{r'})$
but, as in~\cite{Ipina2014}, we have relaxed here the assumption on
the Poisson distribution.  Namely, the assumption that the molecule
numbers follow a Poisson distribution is valid if the species
diffuses, but in the case with $D_S=0$, $N^{(b)}$ and $N^{(S)}$ are
correlated. In such a case, we assume that they follow a binomial
distribution (with $N^{(S)}+N^{(b)}=N_{ST}$ constant), so that
$\langle \delta c^{(S)}(\vec{r},t) \delta c^{(b)}(\vec{r'},t)\rangle =
-\langle {\delta c^{(b)}}(\vec{r},t){\delta
  c^{(b)}}(\vec{r'},t)\rangle=-\langle
N^{(b)}\rangle(1-f_b)\delta(\vec{r}-\vec{r'})2^{3/2}/V_{obs}$, where
$f_b=\langle N^{(b)}\rangle/N_{ST}$ is the fraction of bound molecules
in $V_{obs}$~\cite{Ipina2014}. Including the initial correlation
matrix in Eq.~(\ref{Go_krichevsky}) provides a unifying notation which
embraces all the situations analyzed in our
Letter. Eq.~(\ref{Go_krichevsky}) generally does not have an
analitycal solution, but in \cite{Ipina2013} it was shown that there
exist two limits in which the three eigenvalue branches of the system
can be expressed as $\lambda_i = -\nu_i - D_iq^2$. These limits are
the \textit{fast reaction} limit, (\textit{fr}), in which reactions
take place on a much faster time scale than diffusion in $V_{obs}$,
and the \textit{fast diffusion} limit, (\textit{fd}), in which the
opposite relationship between the time scales holds. In both these
limits it is:
\begin{equation}
\begin{aligned}
G^{(s)}(\tau) &=& \sum_{i=1}^3 G_i^{(s)}(\tau)=\sum_{i=1}^3 \, \frac{{G}_{oi}^{(s)}}{\left(1 +
\frac{\tau}{\tau_{Di}}\right)^{3/2}} e^{-\nu_i \tau},\quad\, \label{eq:G_i}
\end{aligned}
\end{equation}
where $\tau_{Di}={w_r^2}/{4D_i}$ and the weights, ${G}_{oi}^{(s)}$,
are linear combinations of the covariances between the stochastic
variables that satisfy $G_o^{(s)}\equiv\sum_i
{G}_{oi}^{(s)}={\operatorname{var}(N^{(s)})}$. Deriving from
Eq.~(\ref{Go_krichevsky}) the weights, $G_{oi}^{(f)}$ and
$G_{oi}^{(b)}$, in both limits as done in~\cite{Ipina2013} but including
the matrix of initial correlations as done in~\cite{Ipina2014} we
obtain the expressions listed in table~\ref{Table:expressions_ACF}. In
this table we show the values obtained when $D_S=0$ or $D_S \neq 0$ in
the fast diffusion and the fast reaction limits. From these values,
the relative weights, $W_i^{(s)}= {G_{oi}^{(s)}}/G_o^{(s)}$, $s=f,S$,
can be derived.

\begin{table}[H]
\begin{center}
\resizebox{\linewidth}{!}{%
\begin{tabular}{cc|c|c|c|c|c|c|c|c|c|c|c|c|}
\cline{3-14}
\multicolumn{2}{c|}{} & \multicolumn{4}{ c| }{} & \multicolumn{4}{ c |}{}  & \multicolumn{4}{ c| }{}\\
\multicolumn{2}{c|}{\multirow{4}{*}{}} & \multicolumn{4}{ c| }{$G_1(\tau)$} & \multicolumn{4}{ c |}{$G_2(\tau)$}  & \multicolumn{4}{ c| }{$G_3(\tau)$}\\
\multicolumn{2}{c|}{} & \multicolumn{4}{ c| }{} & \multicolumn{4}{ c |}{}  & \multicolumn{4}{ c| }{}\\
\cline{3-14}
& & & & & & & & & & & & &\\
& & $G_{o1}^{(f)}$ & $G_{o1}^{(b)}$ & $\tau_{D1}$ & $\nu_1$ & $G_{o2}^{(f)}$ & $G_{o2}^{(b)}$ & $\tau_{D2}$ & $\nu_2$ & $G_{o3}^{(f)}$ & $G_{o3}^{(b)}$ & $\tau_{D3}$ & $\nu_3$ \\
& & & & & & & & & & & & &\\
\hline
\multicolumn{1}{|c|}{} & & & & & & & & & & & & &\\
\multicolumn{1}{|c|}{\multirow{3}{*}{$D_S \neq 0$}} & \textit{fr} 
& $0$ & $f_b \langle N^{(b)}\rangle$ & $\frac{w_r^2}{4D_S}$ & $0$ 
& $\frac{1}{\left(1+\beta\right)}\langle N^{(f)}\rangle$ & $\frac{\left(1-f_b\right)\beta}{\left(1+\beta\right)}\langle N^{(b)}\rangle$ & $\frac{w_r^2}{4D_{coll}}$ & $0$ 
& $\frac{\beta}{\left(1+\beta\right)}\langle N^{(f)}\rangle$ & $\frac{\left(1-f_b\right)}{\left(1+\beta\right)}\langle N^{(b)}\rangle$ & $\frac{w_r^2}{4D_{ef_2}}$ & $\frac{k_{off}(1+\beta)}{1-f_b}$\\
\multicolumn{1}{|c|}{} & & & & & & & & & & & & & \\
\multicolumn{1}{|c|}{} & \textit{fd} 
& $0$ & $f_b\langle N^{(b)}\rangle$ & $\frac{w_r^2}{4D_S}$ & $0$ 
& $0$ & $(1-f_b)\langle N^{(b)}\rangle$ & $\frac{w_r^2}{4D_S}$ &  $\frac{k_{off}(1+\beta)}{1-f_b}$ 
& $\langle N^{(f)}\rangle$ & $0$ & $\frac{w_r^2}{4D_f}$ & $0$\\
\multicolumn{1}{|c|}{} & & & & & & & & & & & & &\\
\hline
\multicolumn{1}{|c|}{} & & & & & & & & & & & & &\\
\multicolumn{1}{|c|}{\multirow{3}{*}{$D_S = 0$}} & \textit{fr} 
& $0$ & $0$ & $\frac{w_r^2}{4D_S}$ & $0$ 
& $\frac{1}{\left(1+\beta\right)}\langle N^{(f)}\rangle$ & $\frac{\left(1-f_b\right)\beta}{\left(1+\beta\right)}\langle N^{(b)}\rangle$ & $\frac{w_r^2}{4D_{coll}}$ & $0$ 
& $\frac{\beta}{\left(1+\beta\right)}\langle N^{(f)}\rangle$ & $\frac{\left(1-f_b\right)}{\left(1+\beta\right)}\langle N^{(b)}\rangle$ & $\frac{w_r^2}{4D_{ef_2}}$ & $\frac{k_{off}(1+\beta)}{1-f_b}$ \\
\multicolumn{1}{|c|}{} & & & & & & & & & & & & &\\
\multicolumn{1}{|c|}{} & \textit{fd} & $0$& $0$ & $\frac{w_r^2}{4D_S}$ & $0$ 
& $0$ & $(1-f_b)\langle N^{(b)}\rangle$ & $\frac{w_r^2}{4D_S}$ & $\frac{k_{off}}{1-f_b}$ 
& $\langle N^{(f)}\rangle$ & $0$ & $\frac{w_r^2}{4D_f}$ & $0$\\
\multicolumn{1}{|c|}{} & & & & & & & & & & & & & \\
\hline
\end{tabular}
}
\caption{Expressions for the weights, $G_{oi}^{(f)}$ and $G_{oi}^{(b)}$, in the cases with $D_S=0$ or $D_S \neq 0$ and in the fast diffusion (\textit{fd}) or the fast reaction (\textit{fr}) limits, where $f_b=\frac{\langle N^{(b)}\rangle}{N_{ST}}$, $D_{coll}=\frac{D_f + \beta D_S}{1+\beta}$, $D_{ef_2}=\frac{\beta D_f +D_S}{1+\beta}$ and $\beta=\frac{{\langle N^{(S)}\rangle}^2}{K_DN_{ST}V_{obs}}$. In the Letter we assume that, in the fast reaction limit, $\nu_3$ is so large that the corresponding weight is almost negligible when the slow diffusion timescale associated to $D_{ef_2}$ is reached so that it is not   necessary to include the latter in the description. The opposite situation is assumed to hold in the fast diffusion limit between the timescales associated to $D_S$ and $\nu_2$.}
\label{Table:expressions_ACF}
\end{center}
\end{table}

\section{Numerical Simulations}
We performed stochastic numerical simulations of the
reaction-diffusion system considered with a Gillespie-like
algorithm~\cite{Gillespie, Ipina2013}. To compute $N^{(s)}(t)$ we
either worked as with the analytic computations, namely, we computed
$N^{(s)}= \sum_i I(\vec{r}_i)N^{(s)}(\vec{r}_i,t)$ where
$N^{(s)}(\vec{r}_i,t)$ is the number of molecules of species $s$ in a
volume of size, $d\vec{r}_i=1.25\times 10^{-4}\mu m^3$, centered at
$\vec{r}_i$, or we counted all those inside a cube of size $0.016\mu
m^3$ with the same weight ({\it i.e.}, we computed $N^{(s)}= \sum_i
I(\vec{r}_i)N^{(s)}(\vec{r}_i,t)$ with $I=1$ for $ \vert x \vert \le 0.25\mu m
$, $ \vert y \vert \le 0.25\mu m$, $ \vert z \vert\le 0.25\mu m$, and $I=0$ otherwise) .
For the system parameters we used the ones listed in the Table which
were derived from an analysis~\cite{Sigaut2014} of FCS experiments
performed in \textit{Drosophila melanogaster} to estimate the
diffusion coefficient of the protein Bicoid~\cite{AbuArish:BJ2010}.

\begin{table}[H]
\begin{center}
\begin{tabular}{c | c | c | c | c }
 & Fig. 1(a) & Fig. 1(b-c) & Fig. 2(a-b) & Fig. 2(c) \\
\hline
\hline
$D_S$ & $[0-5] \mu m^2s^{-1}$ & $[0-5] \mu m^2s^{-1}$ &  $0 \mu m^2s^{-1}$ & $ 0 \mu m^2s^{-1}$\\
$D_f$ & $19 \mu m^2s^{-1}$ &  $19 \mu m^2s^{-1}$ &  $19 \mu m^2s^{-1}$ &  $19 \mu m^2s^{-1}$\\
$k_{off}$ & $[0.5 - 1 - 5] s^{-1}$ & $400 s^{-1}$ & $400 s^{-1}$ &  $0.5 s^{-1}$ \\
$K_D$ & $1.92 nM$ & $0.2496 \mu M$ & $0.2496 \mu M$ & $1.92 nM$ \\
$[S]$ & $22.1 nM$ & $2.87 \mu M$ & $2.87 \mu M$ & $22.1 nM$ \\
$[P^{(f)}]$ & $59.1 nM$ & $7.68 \mu M$ & $7.68 \mu M$ & $59.1 nM$  \\
$[P^{(b)}]$ & $679.3 nM$ &  $88.31 \mu M$ &  $88.31 \mu M$ & $679.3 nM$ \\
$T_{obs}$ & $100s$ & $100s$ & $100s$ & $100s$ \\
$I(\vec{r})$ & Gaussian & Gaussian & Cubic & Cubic\\
$V_{obs}$ & $0.053 \mu m^3$ & $0.053 \mu m^3$ & $0.016 \mu m^3$ & $0.016\mu m^3$ \\
\hline
\end{tabular}
\caption{Parameters of the simulations.}
\end{center}
\end{table}

\bibliography{supporting_file}